# Magnetic anisotropy in FeSb studied by $^{57}$Fe Mössbauer spectroscopy

K. Komędera[1], A. K. Jasek[1], A. Błachowski[1], K. Ruebenbauer[1*], and A. Krztoń-Maziopa[2]

[1]Mössbauer Spectroscopy Division, Institute of Physics, Pedagogical University
ul. Podchorążych 2, PL-30-084 Kraków, Poland

[2]Warsaw University of Technology, Faculty of Chemistry
ul. Noakowskiego 3, PL-00-664 Warsaw, Poland

[*]Corresponding author: sfrueben@cyf-kr.edu.pl

**Abstract**

The $Fe_{1+x}Sb$ compound has been synthesized close to stoichiometry with x = 0.023(8). The compound was investigated by $^{57}$Fe Mössbauer spectroscopy in the temperature range 4.2 – 300 K. The antiferromagnetic ordering temperature was found as 232 K i.e. much higher than for the less stoichiometric material. Regular iron was found to occupy two different positions in proportion 2:1. They differ by the electric quadrupole coupling constants and both of them exhibit extremely anisotropic electric field gradient tensor (EFG) with the asymmetry parameter $\eta \approx 1$. The negative component of both EFGs is aligned with the *c*-axis of the hexagonal unit cell, while the positive component is aligned with the <120> direction. Hence, a model describing deviation from the NiAs $P6_3/mmc$ symmetry group within Fe-planes has been proposed. Spectra in the magnetically ordered state could be explained by introduction of the incommensurate spin spirals propagating through the iron atoms in the direction of the *c*-axis with a complex pattern of the hyperfine magnetic fields distributed within *a-b* plane. Hyperfine magnetic field pattern of spirals due to major regular iron is smoothed by the spin polarized itinerant electrons, while the minor regular iron exhibits hyperfine field pattern characteristic of the highly covalent bonds to the adjacent antimony atoms. The excess interstitial iron orders magnetically at the same temperature as the regular iron, and magnetic moments of these atoms are likely to form two-dimensional spin glass with moments lying in the *a-b* plane. The upturn of the hyperfine field for minor regular iron and interstitial iron is observed below 80 K. Magneto-elastic effects are smaller than for FeAs, however the recoilless fraction increases significantly upon transition to the magnetically ordered state.

# 1. Introduction

Iron antimony binary system is characterized at low pressure and temperature by some substitutional solubility of the antimony in α-Fe, presence of the non-stoichiometric FeSb metallic compound with the excess iron, highly stoichiometric $FeSb_2$ intermetallic compound and α-Sb with virtually none iron dissolved in [1, 2]. The iron mono antimonide crystallizes in the hexagonal structure within $P6_3/mmc$ group in similarity to NiAs [1, 3]. Antimony forms almost perfect hexagonal lattice with two layers along the *c*-axis of the chemical unit cell. These layers are mutually shifted once versus another in the *a-b* plane to conform to the close packing conditions. They are interlaced with the fully occupied hexagonal sheets of iron, the latter having all sites equivalent one to another from the chemical point of view – regular iron. There are two interstitial positions per chemical cell (one per chemical formula) called double tetrahedral interstitials $DTI_1$ and $DTI_2$, respectively. They are accessible to the excess iron atoms with equal probabilities for each of them [4]. The compound is unstable very close to stoichiometry and hence, some interstitial iron is always present within relatively large range of concentration. Hence, the real chemical formula takes on the form $Fe_{1+x}Sb$. The range of the parameter x is reported as $0.08 < x < 0.38$ [2]. It seems that excess iron is distributed randomly over interstitials at least for small departures from stoichiometry. It was found that iron diffusivity is much higher than antimony diffusivity and that iron diffuses via interstitials $DTI_1$ and $DTI_2$ following the chain regular – interstitial – regular [4].

There is a small net magnetic moment per iron atom leading to the magnetic ordering at low temperatures. The ordering temperature strongly depends on stoichiometry and drops with the increase of the excess iron concentration [5-7]. The highest reported ordering temperature amounts to 211 K for $x = 0.13$ [8, 7]. Magnetic moments of iron order in the *a-b* plane in an antiferromagnetic triangular fashion [6]. Neither magnetic moment nor the hyperfine magnetic field is associated with the antimony. However some transferred magnetic hyperfine field was observed on tin substituting antimony [9]. The excess iron is characterized by larger magnetic moment and it orders magnetically as well with moments being perpendicular to the *c*-axis. Mictomagnetic clusters are formed for higher concentration of the interstitial iron, while for the low concentration one observes spin glass [10, 11]. There are controversies concerned with the ordering temperature of the excess iron. Some reports show that this temperature is lower than the ordering temperature of the regular iron [10, 11].

The compound $Fe_{1+x}Sb$ has been investigated previously by the Mössbauer spectroscopy [4, 5, 7-12]. Iron Mössbauer spectra exhibit at least two iron sites even above the magnetic ordering temperature. Such feature has been interpreted in terms of the distinctly different environment for regular and interstitial iron. One has to note as well, that regular iron is octahedrally coordinated by antimony, and one can expect competing contributions from the direction dependent covalent bonds and more isotropic metallic bonds in some similarity to the FeAs despite different crystal structure [13].

Hence, it is important to look more carefully at this system, in particular for the single-phase samples being as close to stoichiometry as possible. Additionally iron pnictides seem interesting compounds as iron-based superconductivity [14] is generated within iron-pnictogen [15, 16] or iron-chalcogen [17, 18] sheets.

## 2. Experimental

Ceramic pellets of iron antimonide material were prepared by solid state reaction technique. To this end the stoichiometric amounts of high purity (at least 99.99 %, Alfa Aesar) powders of iron and antimony were homogenized together under argon atmosphere, pressed into pellets and sealed in evacuated quartz ampoules. Next the ampoules were heated to 1060 °C and annealed over 5 h, afterwards the furnace was slowly (5 °C/h) cooled down to 600 °C and finally to room temperature over next two hours. The pre-synthesized material was then powdered in an inert atmosphere, pressed again into pellets, sealed in evacuated quartz ampoules and re-annealed at 800 °C over 60 h followed by further thermal treatment at 400 °C for another 48 h.

Phase purity of the FeSb sample was characterized by powder X-ray diffraction method. Measurements were performed at room temperature with D8 Advance Bruker AXS diffractometer with $Cu-K\alpha_{1,2}$ (1.5406 Å) radiation. For the measurements a piece of the ceramic pellet was ground into powder under inert atmosphere and loaded into the low background airtight sample holder to protect the material from oxidation. The X-ray pattern, shown in Figure 1, is consistent with the $P6_3/mmc$ group and the lattice constants were found as $a = 0.406(6)$ nm and $c = 0.513(3)$ nm. The refinements of the crystal structure parameters were performed with FULLPROF program [19] with the use of its internal tables. It is known that lattice constants increase with the departure from stoichiometry, and lattice constants for our sample are equal ($c$) or smaller ($a$) than the smallest constants reported [7].

Elemental composition of the prepared sample was studied using μXRF (micro X-ray fluorescence) spectroscopy (Orbis Micro-XRF Analyzer, EDAX). Measurements were carried out in vacuum, applying white X-ray radiation produced by Rh-tube (35 kV and 500 μA). The X-ray primary beam was focused to a spot of 30 μm diameter. Prior to the measurements an elemental calibration of the instrument has been done using as a standard carefully weighted, homogenized and pressed into pellet mixture of high purity powders of Sb and Fe. The stoichiometry of the prepared material was found to be $Fe_{1.023(8)}Sb_{1.000(2)}$. Hence, highly stoichiometric sample with x = 0.023(8) was obtained.

Mössbauer spectra for 14.41-keV transition in $^{57}$Fe have been collected in standard transmission geometry applying commercial $^{57}$Co(Rh) source kept under ambient pressure and at room temperature. Absorber was made in the powder form mixing 43 mg of FeSb with the $B_4C$ carrier. Absorber thickness amounted to 21.4 mg/cm$^2$ of FeSb with a natural isotopic composition. A Janis Research Co. SVT-400 cryostat was used to maintain the absorber temperature, with the long time accuracy better than 0.01 K (except at 4.2 K, where the accuracy was better than 0.1 K). A RENON MsAa-3 Mössbauer spectrometer equipped with a Kr-filled proportional counter was used to collect spectra in the photo-peak window. Velocity scale of the Mössbauer spectrometer was calibrated by using Michelson-Morley interferometer equipped with the He-Ne laser. Spectral shifts are reported versus ambient pressure and room temperature natural α-Fe. Spectra were fitted within transmission integral approximation by means of the GMFeAs application of Mosgraf-2009 [20].

## 3. Results and discussion

Mössbauer spectra are shown in Figure 2. One can see that magnetic order starts at about 232 K, i.e., much higher than for less stoichiometric samples [7]. Fully developed magnetic spectra at lowest temperatures could be fitted assuming two distinct regular iron sites contributing to the resonant cross-section in proportion 2:1 (called major and minor regular iron) and some additional site with larger hyperfine field contributing 2.1(5) % to the cross-section due to excess interstitial iron. Hence, the chemical formula established by means of the Mössbauer spectroscopy reads as $Fe_{1.021(5)}Sb$ in perfect accordance with the µXRF result. Regular iron sites are best described by spirals of the hyperfine fields propagating along the *c*-axis and probably incommensurate with the respective lattice period. The excess iron is characterized by single hyperfine field larger than the average field of the proposed spirals.

Average magnetic hyperfine fields for all three iron sites and versus temperature are shown in Figure 3, while the total spectral shifts and quadrupole coupling constants for regular iron sites and versus temperature are shown in Figure 4. Even high temperature (paramagnetic) Mössbauer spectra exhibit two different sites in a rough proportion 2:1. It is assumed that the recoilless fraction is isotropic and practically the same for all iron sites (at all temperatures investigated) and sample has random orientation. These two sites differ by the electric quadrupole splitting, while the total shift is practically the same for both of them at high temperature. The site with smaller quadrupole splitting shows two times larger absorption cross-section in comparison with the second site. One can learn more about the behavior of the electric field gradient tensor looking at the magnetically split spectra. For hyperfine fields lying in the *a-b* plane in accordance with the established magnetic structure one can see that the electric field gradient is extremely anisotropic with the asymmetry parameter $\eta = (V_{11} - V_{22})/V_{33} \approx 1$. This tensor has one of the principal components $V_{33}$ oriented along the *c*-axis and negative. Remaining principal components stay in the *a-b* plane with $V_{11} \approx 0$ in accordance with the observed asymmetry. These statements apply to both sites except for the slight difference in the value of the coupling constant. The coupling constant changes also upon transition to the magnetically disordered state – probably due to the magneto-elastic effects in similarity to the case of FeAs [13]. For major iron sites coupling constant decreases while for the minor iron sites increases in the absolute terms. Hence, it is obvious that iron sheets do not conform to the $P6_3/mmc$ symmetry, as for the latter case all regular iron positions are equivalent and the electric field gradient tensor should be axially symmetric with the principal component being aligned with the *c*-axis and $\eta = 0$. On the other hand, the negative value for $V_{33}$ is consistent with the observed ratio $c/a \approx 1.26$ i.e. smaller than $c/a = \sqrt{8/3}$ characterizing the hexagonal close packed structure with vanishing electric field gradient tensor. One of the possible deformations explaining behavior of the spectra is sketched in Figure 5. For this model the positive principal component $V_{22} = -V_{33}$ is aligned with the <120> or equivalent direction. A total spectral shift starts to differ between above sites only at very low temperature indicating differentiation of the electron density between them and showing that dynamics remains practically the same for both of them. Note that interstitial iron has very little influence on the regular iron sites except hyperfine magnetic fields due to the very low concentration of interstitial iron in the sample investigated.

Magnetic hyperfine fields show some distribution correlated with the orientation in the *a-b* plane (see, Figure 6). Fit with discrete set of three hyperfine fields is very bad. Three fields could be present in principle for undisturbed NiAs structure with the first component being due to regular iron with none adjacent interstitial iron, the second component due to the regular iron with one interstitial iron in the vicinity, and finally due to interstitial iron itself. All remaining configurations do not contribute to the spectrum for x ≈ 0.02. A distribution is typical for the spin density wave (SDW) incommensurate with the crystal periodicity in the direction of the SDW propagation [21]. Due to the fact, that one has triangular anti-ferromagnetic order in the *a-b* plane the only plausible propagation direction is the *c*-axis and transversal SDW. Neutron diffraction was unable to find the spin orientation in the *a-b* plane [5, 6]. Therefore one can conclude that SDW is likely to be of the spiral type. On the other hand, it cannot be simply of the circular type, as circular spirals do not lead to the field distribution observed here. Hence, a spiral magnetic order along the *c*-axis probably occurs. A correlation between particular field value and field orientation in the frame of the electric field gradient tensor is a strong support for the hyperfine field spiral structure. Hyperfine fields along spirals are shown in Figure 6. The shape of the spirals was obtained in the same manner as for FeAs [13]. Spirals are likely to be single spirals along the *c*-axis incommensurate with the corresponding lattice period. Spirals due to the minor regular iron sites have vastly different shape from spirals of the major regular iron sites. Note that major sites have two identical spirals shifted in the phase along the *c*-axis, while minor sites generate a single spiral – according to the triangular antiferromagnetic order [6]. Spirals of the major sites are smoothed by the more or less isotropic contribution from the spin polarized itinerant electrons, while spirals of the minor sites show significant direction dependent covalent effects. Such picture is consistent with lower quadrupole coupling constant (in absolute terms) for the major sites due to the enhanced screening by the itinerant charge in comparison with the more covalent minor sites. The highest order terms (fourth order) describing spiral anisotropy in the *a-b* plane conform to the d symmetry of the electrons with uncompensated spins [13].

Interstitial iron orders magnetically practically at the same temperature as iron of both regular sites (Figure 3). The electric field gradient tensor for interstitial iron atoms is likely to be axially symmetric and oriented along the *c*-axis. Very small contribution due to the interstitial iron makes it practically invisible above magnetic transition. For a magnetic region one can fit this contribution in the first order approximation as far as the electric quadrupole interaction is concerned and one obtains positive quadrupole shift $\varepsilon = +0.2(1)$ mm/s and total spectral shift $S = 0.42(5)$ mm/s. Excited nuclear hyperfine states $|\pm\frac{3}{2}\rangle$ are shifted by $\varepsilon$, while corresponding states $|\pm\frac{1}{2}\rangle$ by $-\varepsilon$ in the first order approximation for a nuclear transition from the ground nuclear state with spin $I_g = 1/2$ to the excited nuclear state with spin $I_e = 3/2$ as for a 14.41-keV transition from the ground to the first excited state in $^{57}$Fe. The shift amounts to $\varepsilon = (3/2)A_Q(3\cos^2\beta - 1)$ for axially symmetric electric field gradient. Therefore for $\beta = \pi/2$ one obtains $A_Q = -(2/3)\varepsilon$ with $A_Q$ being the quadrupole coupling constant for the excited nuclear state (see, caption of Figure 4). Here the angle $\beta$ stands for the angle between principal component of the axially symmetric electric field gradient tensor and hyperfine magnetic field acting on the nucleus. The shift $\varepsilon$ has opposite sign to the expected sign of the coupling constant $A_Q$ ($c/a < \sqrt{8/3}$) indicating that the angle $\beta$ is close to $\pi/2$. Hence, one can conclude that the hyperfine field is nearly perpendicular to the *c*-axis and the spin glass is almost two-dimensional as suggested earlier [10, 11].

One can observe that quadrupole splitting and electron density on the iron nuclei are larger for the interstitial iron as compared to the regular iron. This is a consequence of the lowered itinerant electron density for interstitial sites. The average hyperfine fields of spirals are higher close to the ground state in FeSb as compared to FeAs and the magnetic ordering temperature is much higher as well. Some interesting feature appears in particular for the minor regular iron and interstitial iron. Namely, the hyperfine field unexpectedly jumps up between 80 and 4.2 K. Similar phenomenon was observed by Picone and Clark [10, 11]. This phenomenon requires further research and indicates a development of some additional coupling between magnetic moments of the interstitial and more covalent (minor) regular iron.

*3.1. Iron dynamics (recoilless fraction – spectral area)*

A transmission spectrum $P(v)$ is described by the following expression versus Doppler velocity $v$:

$$P(v) = N_0 \left\{ 1 - \frac{f_s}{\lambda} + \frac{f_s}{\lambda} \int_{-\infty}^{+\infty} d\omega \, \rho(\omega - v) \exp[-\sigma(\omega)] \right\}.$$

(1)

This description is valid for a random, i.e., unpolarizing absorber. The symbol $N_0$ denotes the average number of counts per data channel in the folded spectrum and far off the resonance - baseline. The symbol $0 < f_s < 1$ denotes recoilless fraction of the source. The parameter $\lambda > 1$ accounts for the counts different than due to the resonant line in the photo-peak window, while still in the single channel analyzer window. The emission profile for single line resonantly thin and unpolarized source takes on the form:

$$\rho(\omega - v) = [\Gamma_s / (2\pi)] [(\Gamma_s / 2)^2 + (\omega - v)^2]^{-1}.$$

(2)

The symbol $\Gamma_s > 0$ stands for the intrinsic source line width. The absorption profile $\sigma(\omega)$ is a sum of the following sub-profiles:

$$\sigma(\omega) = t_A \sum_n C_n \left[ \frac{(\Gamma_n / 2)^2}{(\Gamma_n / 2)^2 + (\omega - \omega_n)^2} \right].$$

(3)

The summation goes over all lines with $C_n > 0$ denoting contribution due to the particular line having position at $\omega_n$ and of the line width amounting to $\Gamma_n > 0$. Contributions $C_n$ satisfy the condition $\sum_n C_n = 1$, i.e., they are normalized to unity. Integration of the expression (3) over all energies leads to:

$$A = \int_{-\infty}^{+\infty} d\omega \, \sigma(\omega) = \tfrac{1}{2} \pi \, t_A \sum_n C_n \Gamma_n = \tfrac{1}{2} \pi \, t_A \langle \Gamma \rangle \text{ as } \sum_n C_n \Gamma_n = \langle \Gamma \rangle.$$

(4)

The symbol $\langle \Gamma \rangle$ stands for the weighted average line width within the absorber. On the other hand, a dimensionless resonant absorber thickness $t_A$ takes on the following form for flat, homogeneous absorber with the surface being perpendicular to the beam axis (for collimated beam) $t_A = n_0 \sigma_0 d \langle f \rangle (\Gamma_0 / \langle \Gamma \rangle)$. The symbol $n_0$ stands for the number of resonant nuclei per unit volume, $\sigma_0$ denotes resonant cross-section for absorption, $d$ is the absorber thickness along the beam axis, $0 < \langle f \rangle < 1$ stands for the average recoilless fraction along the beam axis, while $\Gamma_0 > 0$ is the natural line width. In principle, parameters $\sigma_0$ and $\Gamma_0$ could slightly vary from one to another site due to the variation of the internal conversion coefficient caused by various electron densities in the vicinity of the resonant nucleus. However, product $\sigma_0 \Gamma_0$ remains constant. Hence, one obtains $A = \frac{1}{2}\pi n_0 \sigma_0 d \, \Gamma_0 \langle f \rangle$, i.e., the integrated absorption profile is proportional to the average recoilless fraction along the beam direction regardless of the absorber resonant thickness. Upon having selected some reference point (e.g. at some pre-selected temperature of the absorber) one gets:

$$\frac{A}{A_0} = \frac{\frac{1}{2}\pi n_0 \sigma_0 d \, \Gamma_0 \langle f \rangle}{\frac{1}{2}\pi n_0 \sigma_0 d \, \Gamma_0 \langle f_0 \rangle} = \frac{\langle f \rangle}{\langle f_0 \rangle} = \frac{t_A \langle \Gamma \rangle}{t_A^{(0)} \langle \Gamma_0 \rangle}.$$

(5)

Here the index 0 ($\langle f_0 \rangle, \langle \Gamma_0 \rangle, t_A^{(0)}$) refers to the pre-selected reference point. Above approach is valid provided everything remains the same for all spectra within a series except some scalar thermodynamic variable of the absorber e.g. temperature.

The ratio $\langle f \rangle / \langle f_0 \rangle$ for FeSb is plotted versus temperature $T$ in Figure 7. The reference point is chosen at 4.2 K, i.e., close to the ground state of the system and corresponding to the lowest temperature spectrum obtained. Dashed vertical line marks magnetic ordering temperature. Some upturn is clearly visible at magnetic ordering and it is impossible to fit these data with single Debye temperature. On the other hand, one can fit Debye model up to about 220 K obtaining the following Debye temperature $\Theta_F = 356(22)$ K. Hence, the reference recoilless fraction could be calculated back as $\langle f_0 \rangle \approx \langle f(T = 4.2 \text{ K}) \rangle \approx \langle f(T = 0 \text{ K}) \rangle \approx 0.89$. One can estimate recoilless fraction at 300 K as $\langle f \rangle \approx 0.63$ using corrected vertical scale of Figure 7. The "jump" is definitely smaller than for FeAs [13], but still visible. The reasons for smaller "jump" are as follows: 1. magnetic ordering temperature is much higher than for FeAs, 2. metalloid is much heavier with less strongly coupled external electrons leading to the more metallic behavior and therefore less covalent bonds between metalloid and iron.

Inset of Figure 7 shows relative spectral area versus temperature calculated as:

$$\text{RSA} = C^{-1} \sum_{k=1}^{C} \left[ (N_0 - N_k)/N_0 \right].$$

(6)

Here the symbol $C$ stands for the number of data channels for the folded spectrum, the symbol $N_0$ denotes the number of counts per channel far-off the resonance (baseline), while the symbol $N_k$ stands for the number of counts in the $k$-th channel [13]. The increase of RSA correlated with the magnetic ordering is clearly seen. The "jump" on the recoilless fraction is definitely smaller than corresponding RSA "jump" due to the saturation effects in the absorber accounted for by the transmission integral approach. Therefore it is important to use proper transmission integral approach while comparing spectral areas in search for the recoilless fraction variation. The "jump" seems to be somewhat smaller than corresponding effect in FeAs [13] in accordance with the "jump" of the recoilless fraction. Hence, one can conclude that lattice hardens upon magnetic ordering, and this feature seems to be common to the iron-pnictogen bonds [13]. There is practically no change in the total shift at the onset of magnetic order. Hence, the second order Doppler shift (SOD) remains unaffected by the magnetic transition. This is an indication that the lattice hardening is due to the change (suppression) of the low frequency phonon modes projected on iron. Additionally, SOD saturates quite rapidly with increasing temperature at the classical values insensitive to the solid state environment. Hence, for relatively high temperature magnetic transition to the magnetically disordered state SOD is insensitive to the changes in the lattice hardness.

## 4. Conclusions

The compound FeSb has been obtained close to stoichiometry with very small amount of the excess interstitial iron and highest observed up to now magnetic ordering temperature of about 232 K.

The symmetry within iron planes is lowered in comparison with the symmetry expected for the P6$_3$/mmc group. Regular iron is located on two different sites in proportion 2:1.

Hyperfine fields on iron are restricted to the *a-b* plane and are likely to form spirals on the regular sites propagating along the *c*-axis and incommensurate with the lattice period along this direction. Spirals are very similar to spirals observed in FeAs despite different crystal symmetry [13]. The reason for that is that these two compounds have octahedral coordination of iron by pnictogen.

**Acknowledgments**

This work was supported by the National Science Center of Poland, Grant DEC-2011/03/B/ST3/00446. AKM acknowledges the financial support founded by the European Union in the framework of European Social Fund through the Warsaw University of Technology Development Program (CAS/032/POKL).

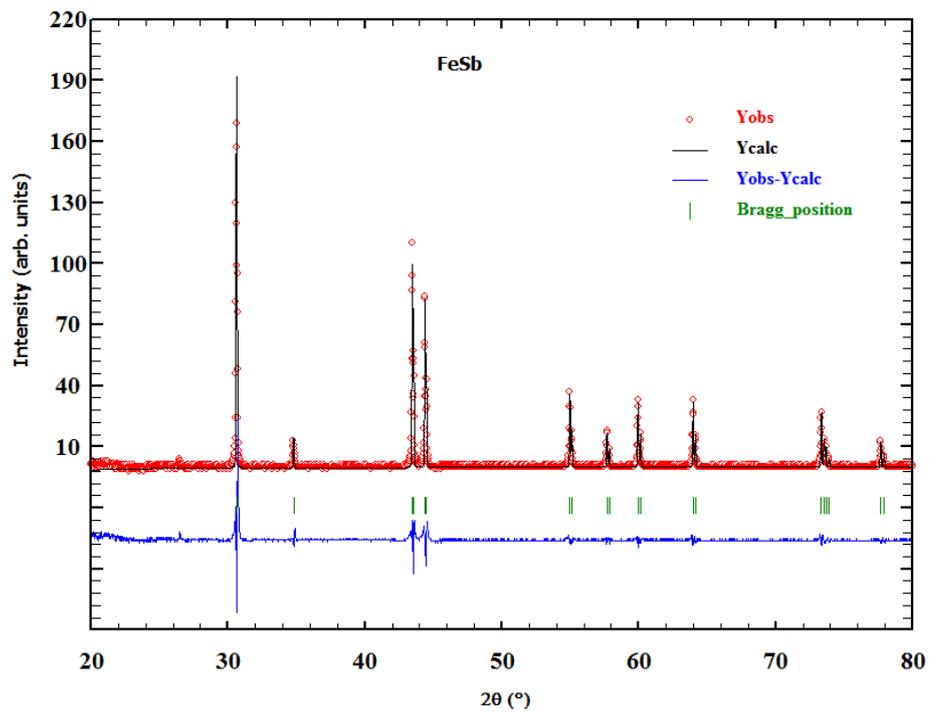

**Figure 1** Powder X-ray diffraction pattern obtained at room temperature using $Cu-K\alpha_{1,2}$ (1.5406 Å) radiation. The symbol 2Θ stands for the scattering angle.

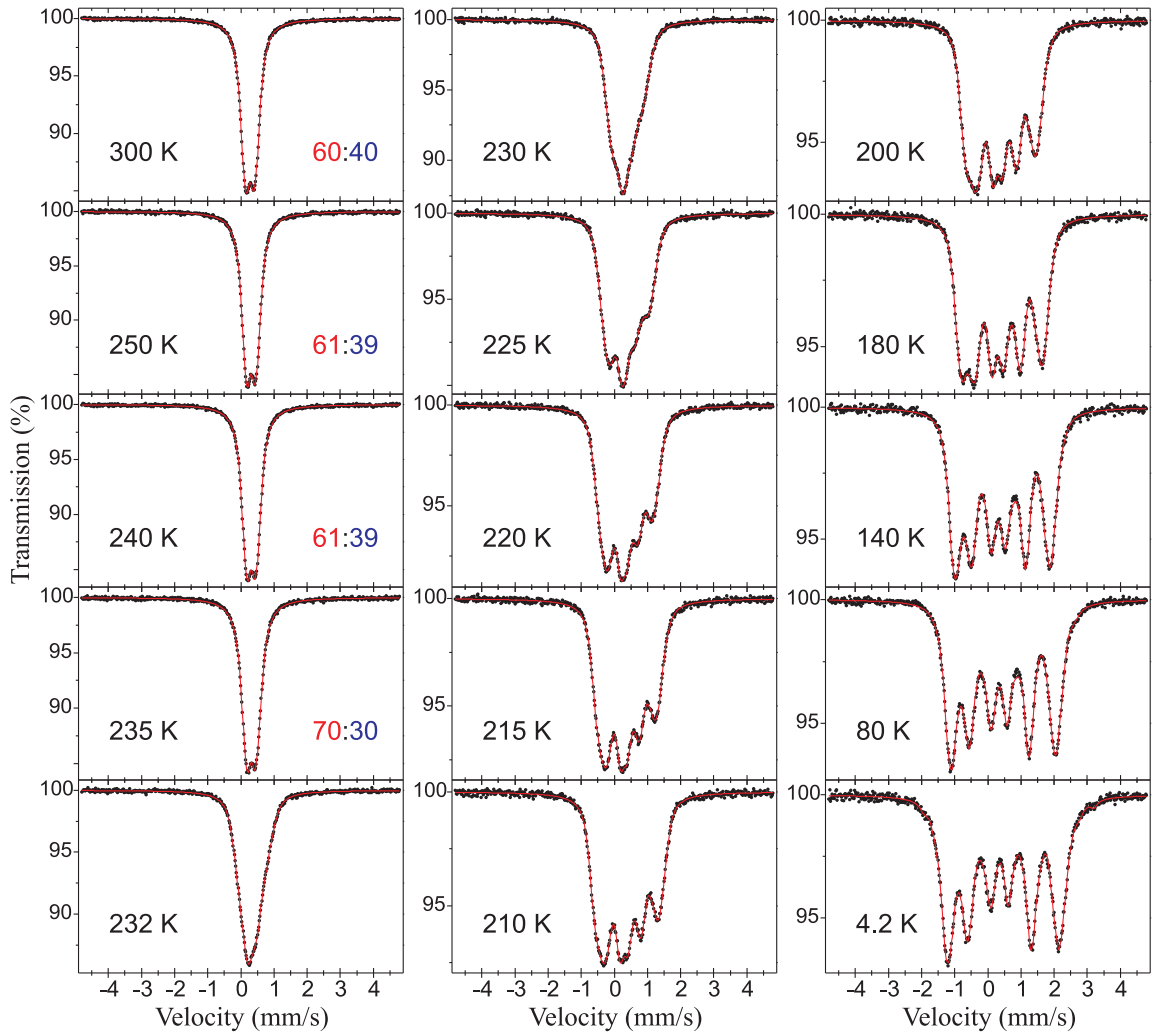

**Figure 2** $^{57}$Fe Mössbauer spectra of FeSb versus temperature. The ratio of the contributions due to the major and minor component is shown as inset for each spectrum collected above magnetic transition (232 K). Interstitial iron is invisible in this region of temperature. For magnetically split spectra one obtains 2:1 ratio of the major to minor regular iron components and 2.1 % contribution to the total cross-section due to interstitial iron.

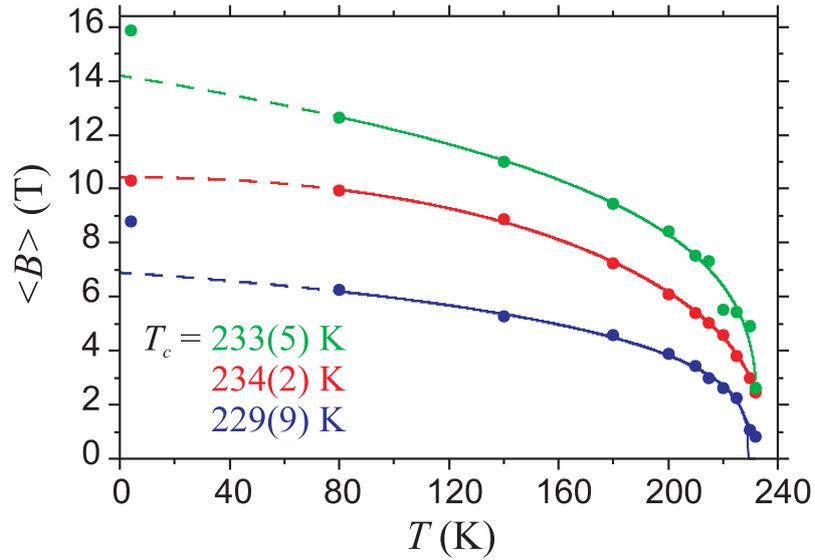

**Figure 3** Average magnetic hyperfine fields versus temperature. The major regular iron is shown in red, the minor regular iron in blue, while a contribution due to interstitial iron in green. The symbol $T_c$ denotes respective transition temperatures being in fact common to all three subsystems considering errors. Experimental points at the lowest temperature (4.2 K) were excluded from fits. Data fitting method is described in Ref. [21].

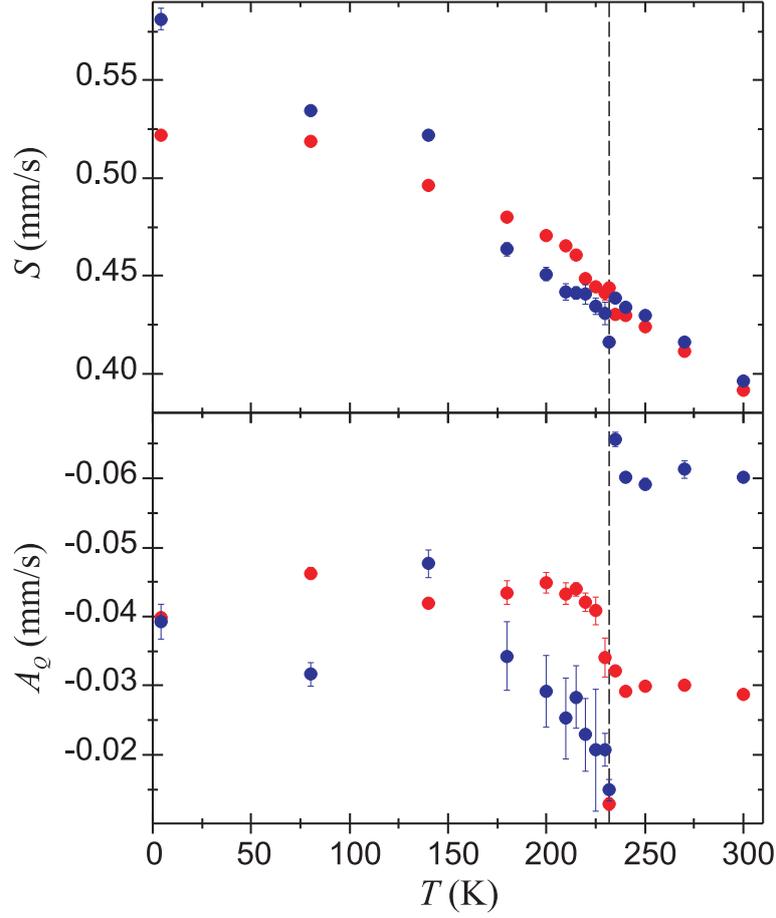

**Figure 4** Total spectral shift $S$ versus room temperature α-Fe and quadrupole coupling constant $A_Q = \left(\dfrac{eQ_e V_{33}}{4I_e(2I_e-1)}\right)\left(\dfrac{c}{E_0}\right)$ versus temperature for regular iron (major – red, minor – blue). The symbol e denotes positive elementary charge, the symbol $Q_e = +0.17$ b [22] stands for the spectroscopic electric quadrupole moment of the first excited state in $^{57}$Fe, while the symbol $I_e = 3/2$ denotes nuclear spin of the above nuclear state. The symbol c stands for the speed of light in vacuum, while the symbol $E_0$ denotes transition energy from the first excited to the ground nuclear state in $^{57}$Fe (14.41-keV). The quadrupole splitting in non-magnetic region amounts to $\Delta = 6|A_Q|\sqrt{1+\eta^2/3}$ with $\eta = 1$. Dashed vertical lines at 232 K represent magnetic ordering temperature.

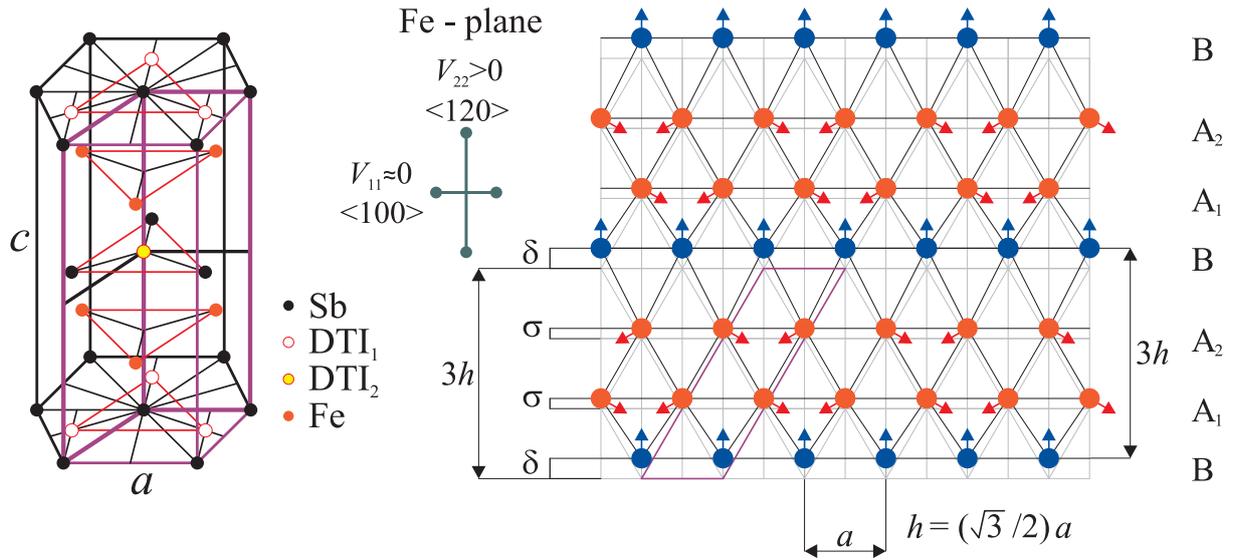

**Figure 5** Crystal structure of the FeSb and proposed modification of the Fe-plane in comparison with the NiAs structure (P6$_3$/mmc). Symbols $a$ and $c$ denote lattice constants, while symbols DTI$_1$ and DTI$_2$ denote interstitial sites accessible to excess iron. The symbol $3h$ stands for the extended period in the Fe-plane along the <120> direction. Deformations are described by the in plane shift parameters δ and σ. Symbols $V_{11}$ and $V_{22}$ denote components of the electric field gradient oriented along <100> and <120> directions, respectively. There are three crystallographic positions of equal probabilities filled by regular iron called B, A$_1$ and A$_2$. Positions A$_1$ and A$_2$ are indistinguishable one from another by the Mössbauer spectroscopy. The same statement applies to equally probable interstitial positions DTI$_1$ and DTI$_2$. Note that the ratio $c/a$ is exaggerated for better visibility.

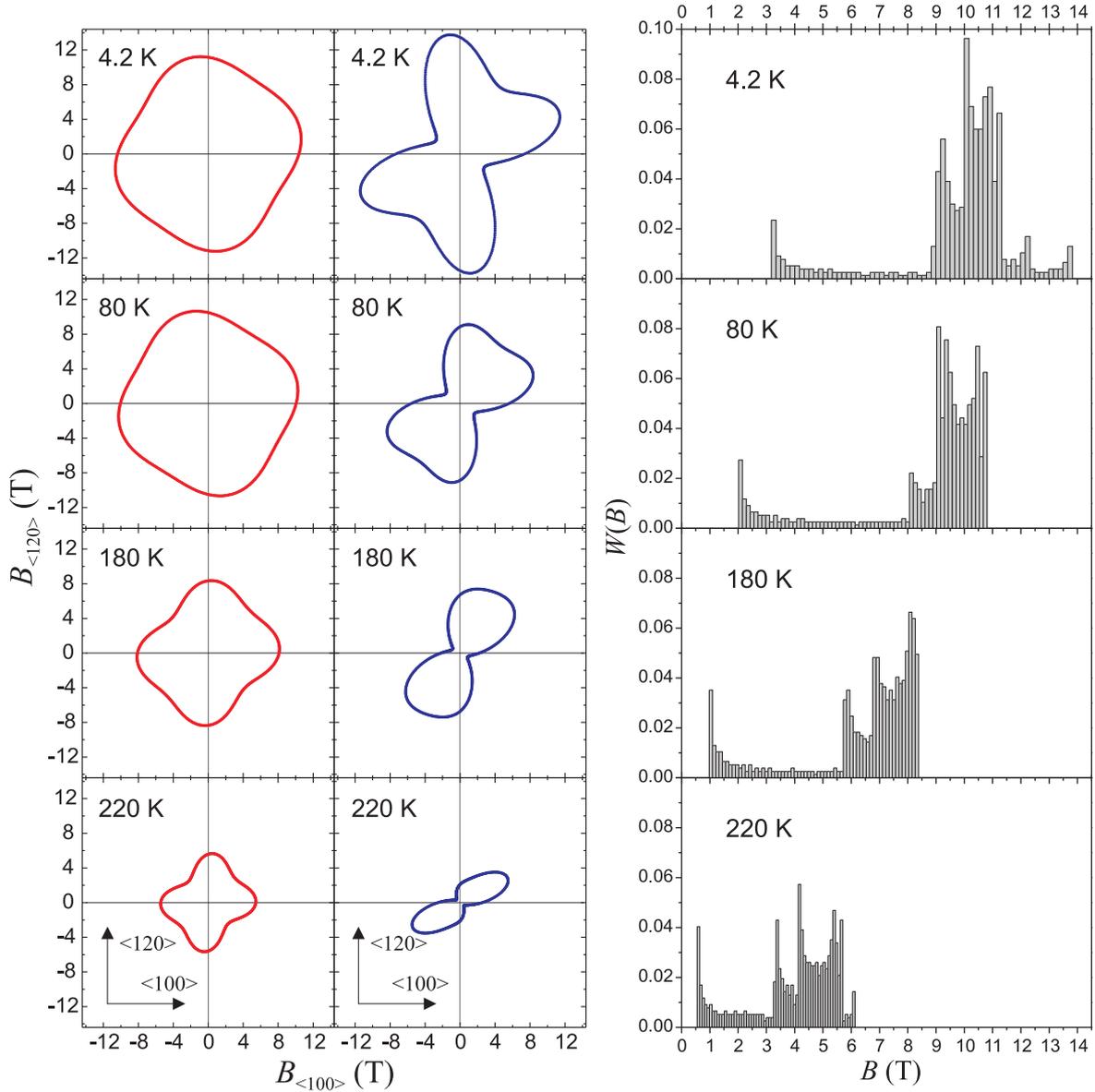

**Figure 6** Magnetic hyperfine field projection on the Fe-plane for spirals propagating along the *c*-axis and incommensurate with the lattice period in this direction. The red color refers to the major regular iron ($A_1$ and $A_2$), while the blue to the minor regular iron (B). The hyperfine field along <100> direction is marked as $B_{<100>}$, while the hyperfine field along <120> direction as $B_{<120>}$. Corresponding total distributions $W(B)$ of the hyperfine field $B$ (averaged over regular sites in proportion 2:1) are shown on the right side. Each of them is normalized to unity.

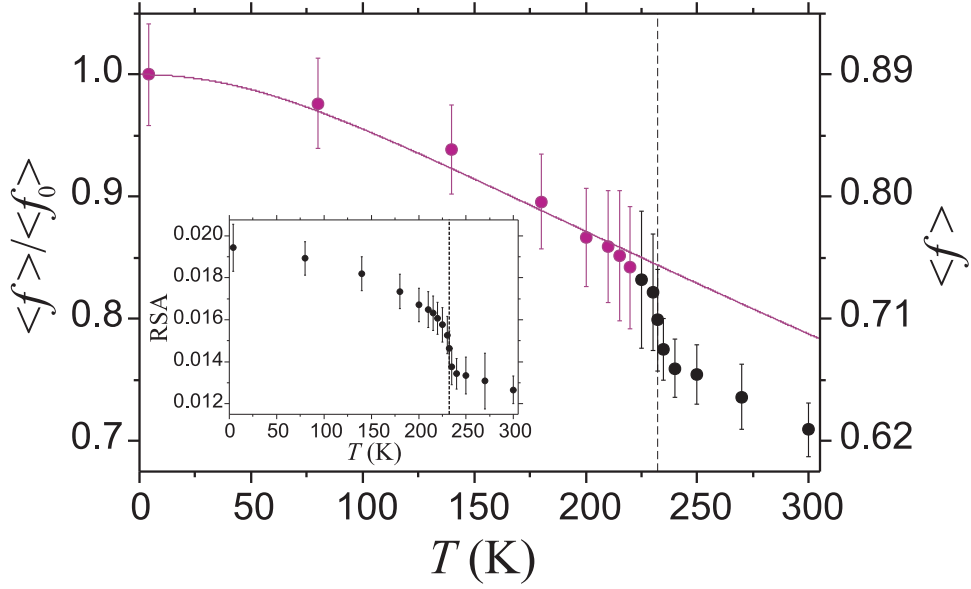

**Figure 7** Ratio of the recoilless fractions $\langle f \rangle / \langle f_0 \rangle$ for FeSb plotted versus temperature. Dashed vertical lines at 232 K mark magnetic ordering temperature. The reference point is taken at 4.2 K. Points shown in purple were used to fit Debye model. Resulting continuous line is calculated within above model with Debye temperature $\Theta_F = 356(22)$ K. The right hand vertical scale shows recoilless fraction $\langle f \rangle$. Inset shows relative spectral area (RSA) calculated basing on the original data.